\documentclass[%
 rsi,
 amsmath,amssymb,
 reprint,%
]{revtex4-1}

\usepackage{graphicx}
\usepackage{dcolumn}
\usepackage{bm}

\usepackage[utf8]{inputenc}
\usepackage[T1]{fontenc}
\usepackage{mathptmx}
\usepackage{etoolbox}
\usepackage{booktabs}
\usepackage{graphicx}

\begin{document}

\preprint{AIP/123-QED}

\title{Correcting the Energy-Dependent Asymmetry in Low-Energy $\mu$SR}
\author{G. Janka}
\email{gianluca.janka@psi.ch}
\affiliation{PSI Center for Neutron and Muon Sciences CNM, 5232 Villigen PSI, Switzerland}

\author{Z. Salman}%
\author{A. Suter}%
\author{T. Prokscha}%
\email{thomas.prokscha@psi.ch}
\affiliation{PSI Center for Neutron and Muon Sciences CNM, 5232 Villigen PSI, Switzerland}

\date{\today}

\begin{abstract}
Low-energy $\mu$SR (LE-$\mu$SR) enables depth-resolved studies of magnetic and electronic properties from the surface into the near-surface region, but the measured transverse-field asymmetry is not an intrinsic constant and depends on implantation energy and beamline conditions. Following an upgrade of the single-muon tagging system at the LEM beamline at PSI in 2023, updated asymmetry calibrations became necessary. Here, we present updated reference measurements that establish the energy-dependent maximum asymmetry using a silver reference and quantify spurious contributions from reflected muons using a nickel reference. In addition, we address systematic reductions of the measured asymmetry arising from incomplete beam--sample overlap by introducing a sample-size-dependent correction factor. This factor is obtained from \texttt{musrSim}/\texttt{Geant4} simulations incorporating an updated electrostatic field map of the sample environment and is benchmarked using implantation-energy scans on SrTiO$_3$ samples of different lateral dimensions. Together, these updated calibrations and the simulation-based overlap correction provide a practical and up-to-date framework for LE-$\mu$SR data correction under current beam conditions and enable quantitative depth-resolved analysis of volume fractions.
\end{abstract}

\maketitle

\section{\label{sec:intro}Introduction}

Low-energy muon spin rotation (LE-$\mu$SR) provides a unique capability to investigate the magnetic and electronic properties of the surface and near-surface regions of materials. At the Low-Energy Muon (LEM) beamline at the Paul Scherrer Institute, a continuous beam of spin-polarized positive muons is moderated and subsequently re-accelerated to kinetic energies up to 30~keV \cite{2008_Prokscha, 2001_Prokscha,2004_Morenzoni}. The final implantation energy at the sample is then set by the combination of the beam transport voltage and the sample bias. Depending on the material, the mean stopping depth ranges from the surface to several hundred nanometers. This enables depth-resolved $\mu$SR measurements with nanometer-scale depth resolution and makes LE-$\mu$SR a powerful technique for studying thin films, multi-layers, surfaces, and interfaces of materials \cite{2022_Blundell, 2024_Amato}.

A defining feature of the LEM beamline is the use of a single-muon tagging system to establish the start time of each $\mu$SR event \cite{1998_Hofer,2015_Khaw}. The muon beam is transmitted through an ultrathin carbon foil, where the emission of secondary electrons provides a precise timing signal. Passage through the foil is accompanied by energy loss and straggling, such that the initially nearly monoenergetic muon beam acquires an energy distribution \cite{2024_Janka}. After passing through the foil, the muons are transported electrostatically toward the sample, where their final implantation energy is determined by a combination of the beam transport voltage and the sample bias. A ring anode located upstream of the sample serves as the final focusing element and defines the beam spot on the sample surface.

The depth resolution of LE-$\mu$SR is intrinsically linked to the electrostatic acceleration and deceleration of the muon beam and requires precise knowledge of the muon energy distribution, beam transport, and sample bias. As a consequence, the fraction of polarized low-energy muons that ultimately reach the sample and contribute to the detected $\mu$SR signal is no longer constant, but depends sensitively on implantation energy, transport settings, and sample geometry. 
In contrast to conventional bulk $\mu$SR experiments, where at sufficiently low transverse magnetic fields the maximum observable asymmetry is essentially a fixed property of the spectrometer, the measured asymmetry in LE-$\mu$SR becomes an energy-dependent quantity.

The asymmetry is the key observable from which we can extract information about the magnetic transition temperature and the magnetic volume fraction in magnetic materials. Similarly, in non-magnetic insulators and semiconductors, the asymmetry can provide information about their electronic properties. In such materials, a fraction of the implanted muons can form muonium (a hydrogen-like bound state of a muon and an electron), while a diamagnetic fraction remains as bare muons upon implantation. These fractions depend on the electronic properties of the host material. Therefore, any modification of the total observable asymmetry directly affects the determination of depth-resolved magnetic and electronic properties. Several instrumental effects contribute to this modification, including muon neutralization in the carbon foil used for single-muon tagging \cite{2024_Janka}, neutralization and backscattering from the sample surface, reflection of low-energy muons in the sample environment, transport- and bias-induced changes of the beam spot, and incomplete overlap between the muon beam and the sample.

Previous studies at LEM have addressed some of these effects and provided corrections under specific beam conditions \cite{2023_Suter}. However, recent upgrades of the LEM beamline, in particular the installation of a new carbon foil with a nominal thickness of 2 nm in the single-muon tagging system \cite{2024_Janka, 2025_Janka}, have significantly altered the beam characteristics and therefore require an updated and comprehensive reassessment of the total asymmetry calibration. In addition, implantation-energy-dependent beam spot effects have so far not been quantitatively incorporated into standard LE-$\mu$SR data analysis.

In this work, we present an updated calibration framework for the depth-dependent total asymmetry at the LEM beamline under current operating conditions. Using dedicated silver and nickel reference measurements, we establish updated energy-dependent reference asymmetries and quantify contributions from reflected muons. We further introduce a simulation-based correction for sample-size-dependent beam-sample overlap, derived from detailed Geant4-based beamline simulations and benchmarked experimentally using SrTiO$_3$ reference measurements. Together, these developments provide a robust and practical procedure for quantitative LE-$\mu$SR analysis, enabling reliable extraction of depth-resolved volume fractions across a wide range of implantation energies and sample geometries.

\section{\label{sec:asymmetry}Asymmetry in LE-$\mu$SR}

\begin{figure}[t!]
\centering
\includegraphics[width=1\columnwidth, trim={0 0 0 0},clip]{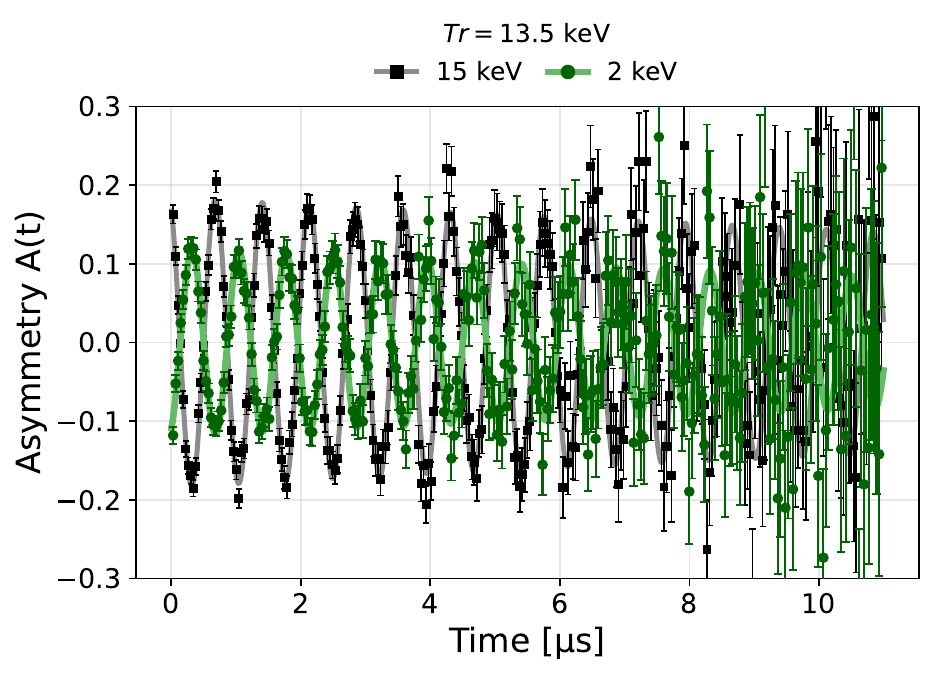}
\caption{\label{fig:STO_example}
Comparison of wTF $\mu$SR time spectra of a $15\times15\,\mathrm{mm}^2$ SrTiO$_3$ sample measured at $200\,\mathrm{K}$ in an applied field of $100\,\mathrm{G}$.
Data were acquired at implantation energies of $2\,\mathrm{keV}$ (green circles) and $15\,\mathrm{keV}$ (black squares) for a transport setting of $13.5\,\mathrm{keV}$. The corresponding fits to Eq.~\ref{eq:asym} are shown in a slightly lighter shade of the respective color. 
For improved visual comparison, the $15\,\mathrm{keV}$ data were taken from the left detector and the $2\,\mathrm{keV}$ data from the right detector, resulting in a relative phase shift of $180^\circ$.}
\end{figure}

Weak transverse-field (wTF) data at LEM are commonly represented in terms of the time-dependent decay asymmetry,
\begin{equation}
A(t)=\frac{N(t)-N_{\mathrm{bg}}}{N_{0}}\,\mathrm{e}^{t/\tau_\mu}- 1,
\end{equation}
where $N(t)$ denotes the accumulated positron counts in a given decay positron detector as a function of time, $N_{0}$ is a normalization factor, $N_{\mathrm{bg}}$ is the constant background contribution, and $\tau_\mu \simeq 2.2$~$\mu$s is the muon lifetime. Physical information is typically extracted by fitting the asymmetry with, for example, an exponentially damped oscillatory function,
\begin{equation}
\label{eq:asym}
A(t)=A_{0}\,\mathrm{e}^{-\lambda t}\cos\!\left(\omega_L t + \phi\right),
\end{equation}
where $A_0$ is the initial decay asymmetry at $t=0$, $\lambda$ is the depolarization rate, $\omega_L$ is the Larmor precession frequency in the applied magnetic field $B$ ($\omega_L = \gamma_\mu B$, with $\gamma_\mu/2\pi = 135.5$~MHz/T), and $\phi$ is a phase offset depending on the position of the decay positron detector with respect to the initial muon spin direction. Additional details on the fitting procedure and polarization functions of $\mu$SR data can be found in Refs.~\cite{1997_Dalmas, 2022_Blundell,2024_Amato}. Representative examples of such fitted spectra are shown in Fig.~\ref{fig:STO_example}. Note that all experimental $\mu$SR data presented here were analyzed using the \texttt{musrfit} software package~\cite{2012_Suter}. 

In bulk $\mu$SR experiments, the dependence of $A_0$ or $\lambda$ on parameters such as temperature, magnetic field, or pressure provides direct insight into the magnetic and electronic properties of the system. In LE-$\mu$SR, the implantation energy introduces an additional degree of freedom, allowing the asymmetry and depolarization rate to be studied as a function of depth below the surface.

In a wTF measurement, where the applied field is smaller than the static internal fields in a magnetic material, a reduction of $A_0$ upon cooling indicates the onset of magnetic order. This loss is attributed to muons stopping in the magnetically ordered regions of the sample. There, the muons experience a fast depolarization due to the large static local fields or large widths of local field distributions, and therefore, do not contribute to the oscillating signal at the Larmor frequency. 
Similarly, in LE-$\mu$SR measurements, a loss of observable asymmetry can arise from muonium formation, in which case the muon spin polarization is dominated by electron Zeeman and hyperfine interactions. In typical cases where the hyperfine coupling is strong (e.g., vacuum-like muonium), the associated precession frequencies are orders of magnitude higher than $\omega_L$ and therefore lie far beyond the time resolution of the LEM spectrometer, which is approximately 5~ns (RMS) \cite{2012_Prokscha}. As a result, muonium formation manifests itself as a reduction or loss of $A_0$, where the observed asymmetry reflects only the diamagnetic (bare) muon fraction.

In many experiments, the quantity of interest is the magnetic volume fraction ($V_M$) or the diamagnetic fraction ($f_\mathrm{dia}$) rather than the absolute asymmetry itself. For SrTiO$_3$ (STO), a non-magnetic insulator in which muonium forms only below approximately 70~K \cite{2014_ZaherSTO}, the diamagnetic fraction as a function of temperature can be obtained from
\begin{equation}
\label{eq:diaT}
f_{\mathrm{dia}}(T)
=
\frac{A_{\mathrm{STO}}(T) - A_{\mathrm{bg}}}
     {A_{\mathrm{STO}}(T=290~\mathrm{K})- A_{\mathrm{bg}}},
\end{equation}
where the asymmetry at room temperature serves as a reference corresponding to all bare muons, and $A_\mathrm{bg}$ denotes a temperature-independent background contribution. This relative normalization approach is valid provided that such a reference state can be experimentally accessed.

If a full diamagnetic fraction (or fully paramagnetic volume fraction) reference state cannot be reached within the experimentally available temperature range (up to $320$~K for a typical cryostat setup at LEM), an alternative normalization scheme is required. In this case, the determination of $V_M$ or $f_\mathrm{dia}$ relies on knowledge of the absolute total asymmetry rather than on relative changes with temperature alone. Since this absolute asymmetry is an instrument-dependent quantity, it must be established through dedicated calibration measurements.

With LE-$\mu$SR, the situation becomes more complex when depth-dependent effects are considered. If $V_M$ or $f_\mathrm{dia}$ varies with implantation energy, an energy-dependent reference asymmetry $A_\mathrm{total}(E)$ is required. For example, the depth-dependent $f_\mathrm{dia}$ can then be expressed as
\begin{equation}
\label{eq:diaE}
f_{\mathrm{dia}}(E)
=
\frac{A_{\mathrm{STO}}(E)-A_\mathrm{bg}(E)}
     {A_{\mathrm{total}}(E)-A_\mathrm{bg}(E)}.
\end{equation}
However, this expression only holds if the sample area is much larger than the muon beam spot. If that is not the case, an additional correction factor $\mathcal{O}(E)$ for the overlap of muons with the sample must be introduced:
\begin{equation}
\label{eq:diaE_full}
f_{\mathrm{dia}}(E)
=
\frac{A_{\mathrm{STO}}(E)-A_\mathrm{bg}(E)}
     {\mathcal{O}(E)\,\bigl[A_{\mathrm{total}}(E)-A_\mathrm{bg}(E)\bigr]}.
\end{equation}

Determining $A_{\mathrm{total}}(E)$, $A_{\mathrm{bg}}(E)$ and $\mathcal{O}(E)$ reliably is therefore a central requirement for quantitative LE-$\mu$SR analysis and necessitates a careful and systematic assessment of all effects that modify the total observable asymmetry. 

\section{\label{sec:effects}Energy Dependence of the Total Asymmetry}

In LE-$\mu$SR, the total observable asymmetry is not a fixed instrument constant. In contrast to bulk $\mu$SR, several beamline- and geometry-related effects modify the fraction of tagged muons that reach the sample and contribute to the detected signal. As a result, the measured initial asymmetry depends not only on the intrinsic sample properties, but also on implantation energy and transport settings (Tr).

\begin{figure}[t!]
\centering
\includegraphics[width=1\columnwidth, trim={0 0 0 0},clip]{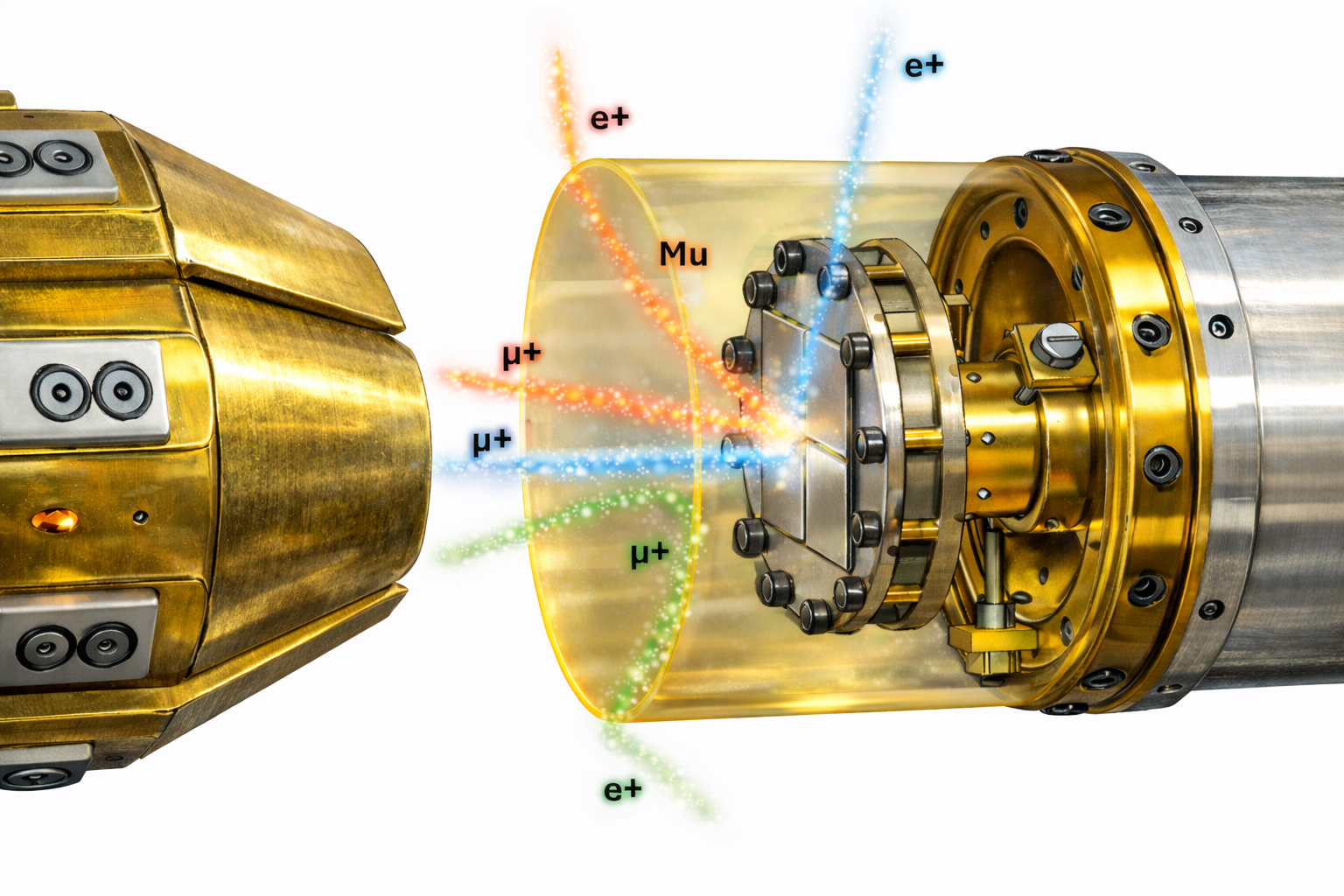}
\caption{\label{fig:effects}
Schematic illustration of the last focusing element (ring anode) and the sample environment at the LEM beamline, highlighting the dominant processes affecting the observable wTF asymmetry in LE-$\mu$SR. Muons stopping in the sample produce the desired $\mu$SR signal (blue). Muons reflected from the sample plate due to insufficient kinetic energy stop in the surrounding radiation shield and can contribute a spurious signal (green). Muons that backscatter from the sample as neutral muonium 
quickly depolarize which leads to a reduction of the observable asymmetry (orange).}
\end{figure}

Fig.~\ref{fig:effects} schematically illustrates the sample environment at the LEM beamline and highlights the dominant processes affecting the observable asymmetry. In the following, the effects of muon backscattering (Fig.~\ref{fig:effects}, orange), reflection (Fig.~\ref{fig:effects}, green), as well as neutralization at the carbon foil are discussed. In addition, we address the effects due to the overlap of the muon beam with the sample and secondary electron emission at the sample.

\subsection{\label{sec:backscattering}Backscattering of Muons}

Upon implantation into a material, a low-energy muon can undergo large-angle scattering in the Coulomb field of a target nucleus, reversing its momentum and exiting the sample. These muons are referred to as backscattered muons and are depicted by the orange trajectory in Fig.~\ref{fig:effects}. In this process, a muon typically loses about two thirds of its initial kinetic energy before reaching the surface, which increases the probability of electron capture, and therefore, it exiting as a neutral muonium \cite{Morenzoni2002NIMPRSB, Biersack1984APA}. 
If the backscattered muons or muonium stop within the sample surroundings (e.g., in the radiation shield), their decay positrons are detected by the LE-$\mu$SR spectrometer.
However, due to their high precession frequencies, muonium events do not contribute to the observable asymmetry and lead to a reduction of the measured signal.
The probability of muonium formation and escape increases strongly at lower implantation energies, leading to a pronounced decrease in the observable asymmetry at shallow depths. 
In contrast, if backscattered muons or muonium escape upstream along the beamline, their decays contribute predominantly within the first $\sim$100~ns of the time spectrum. This results in a reduction of the observable asymmetry; however, unlike the case where they stop in the sample environment, these events do not exhibit a significant energy dependence.

\subsection{\label{sec:reflecting}Reflection of Muons}

Passage through the ultrathin carbon foil introduces energy loss and straggling \cite{2015_Allegrini, 2016_Allegrini}, such that the initially nearly monoenergetic muon beam acquires an energy distribution \cite{2024_Janka, 2023_Suter, 2015_Khaw, 1998_Hofer}. Implantation energies are defined with respect to the most probable muon energy after the foil, while ignoring the low-energy tail. If muons from this tail encounter a decelerating potential at the sample plate that exceeds their kinetic energy, they are reflected before reaching the sample.

These reflected muons, represented by the green trajectories in Fig.~\ref{fig:effects}, predominantly stop in the surrounding radiation shield and can contribute to a spurious transverse-field signal. Unlike backscattering, this effect can lead to an apparent increase in the measured asymmetry, as the reflected muons precess in a well-defined magnetic environment outside the sample, e.g., when stopping in the radiation shield of the cryostat.

\subsection{\label{sec:overlap}Overlap of the Muon Beam with the Sample}

The overlap between the muon beam and the sample area determines the fraction of muons that contribute to the sample-specific $\mu$SR signal. The achievable beam spot size depends on the transport settings \cite{2024_Janka, 2023_Ni} and is further modified by the electrostatic fields applied to the sample plate, which can focus or defocus the incoming beam.

Therefore, for samples comparable in size to the beam spot, the measured asymmetry depends on transport voltage and sample bias. Muons that miss the sample typically stop in the Ni-coated sample plate and therefore do not contribute to the sample's signal, leading to a systematic reduction in the observed asymmetry.

\subsection{\label{sec:electron_emission}Secondary Electron Emission at the Sample}

Impinging muons on the sample can induce the emission of secondary electrons from its surface, analogous to the process occurring at the carbon foil. Under most experimental conditions, these electrons do not affect the $\mu$SR measurement. However, when the sample plate is biased below approximately $-3.6$~kV, secondary electrons emitted from the sample can be accelerated upstream toward the muon detector, overcome the detector’s potential barrier and generate additional spurious trigger signals.

Since the single-muon tagging detector does not operate with 100\% efficiency (approximately 90\% with a clean foil \cite{2025_Janka}), a fraction of muons reaches the sample without producing a valid start signal. If a secondary electron emitted from the sample subsequently generates a delayed trigger, this can be incorrectly associated with the decay of such an untagged muon. In this case, the effective start time occurs after muon implantation, leading to an artificially shortened time interval between start and positron detection. As a result, the oscillatory $\mu$SR signal is the sum of two phase-shifted precession signals, manifesting as a reduction in the observable asymmetry.

This effect appears above a well-defined bias threshold corresponding to higher implantation energies and depends on the electronic properties of the sample. At present, no general correction scheme is available for this effect.

\subsection{\label{sec:neutralization}Neutralization in the Carbon Foil}

When passing through the carbon foil used for single-muon tagging, a fraction of muons neutralizes and forms muonium. These are unaffected by subsequent electrostatic and magnetic fields and are therefore not focused onto the sample. They also depolarize due to the muon-electron hyperfine interaction. Such neutralized muons result in an effective loss of initial polarization.

The neutralization probability depends strongly on the muon kinetic energy. Measurements at LEM have shown that for a transport voltage of 10~keV, the neutralization probability can be up to a factor of two larger than for 15~keV transport \cite{2024_Janka}. This effect constitutes the dominant origin of the transport-setting dependence of the total asymmetry, with lower transport voltages leading to systematically reduced asymmetry.

\section{\label{sec:data}Calibration and Correction of Data}

\subsection{Experimental Input}

\begin{figure}[b!]
\centering
\includegraphics[width=1\columnwidth, trim={0 0 0 0},clip]{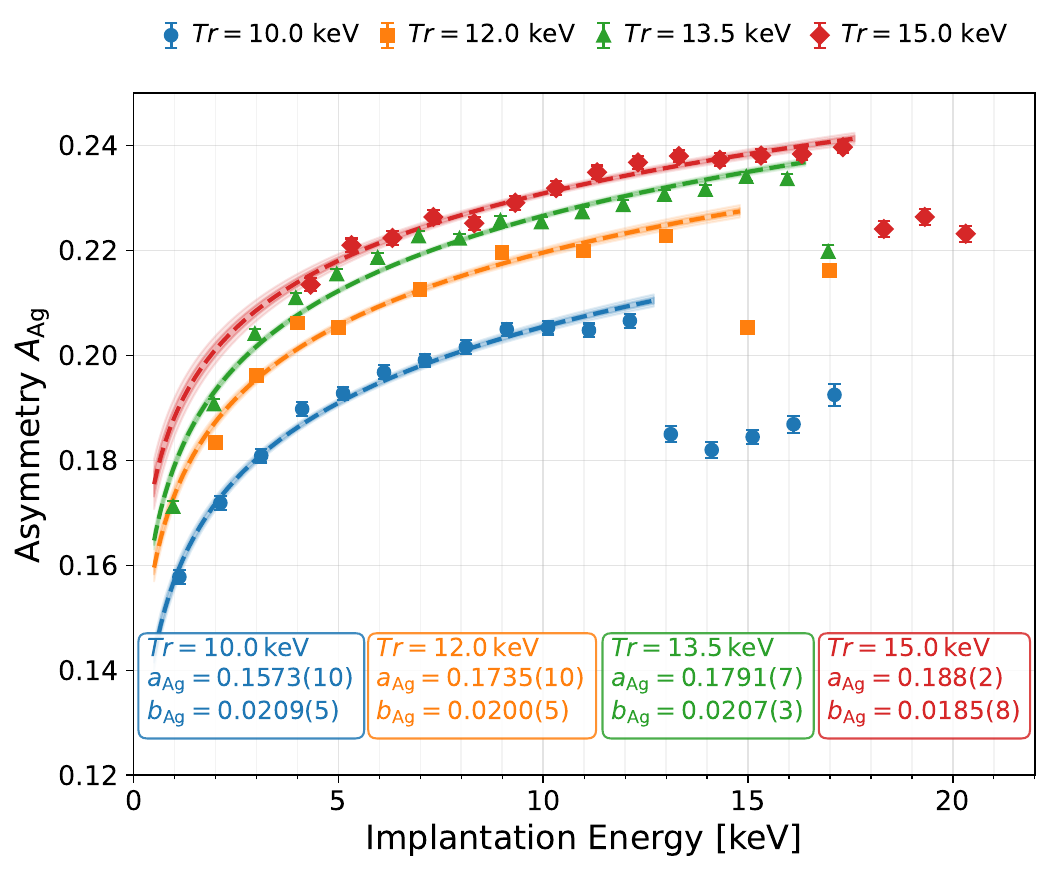}
\caption{\label{fig:backscattering}
Energy-dependent transverse-field asymmetry $A_{\rm Ag}$ measured on a silver-coated sample plate for transport settings of 10 (blue circles), 12 (orange squares), 13.5 (green triangles), and 15~keV (red diamonds). As silver is non-magnetic, the observed energy dependence reflects instrumental effects, including backscattering at shallow implantation depths and transport-dependent losses due to muon neutralization in the carbon foil. The dashed lines represent fits using Eq.~(\ref{eq:asyag}), the fit parameters are summarized in the bottom. 
}
\end{figure}
\begin{figure*}[t!]
\centering
\includegraphics[width=2\columnwidth, trim={0 0 0 0},clip]{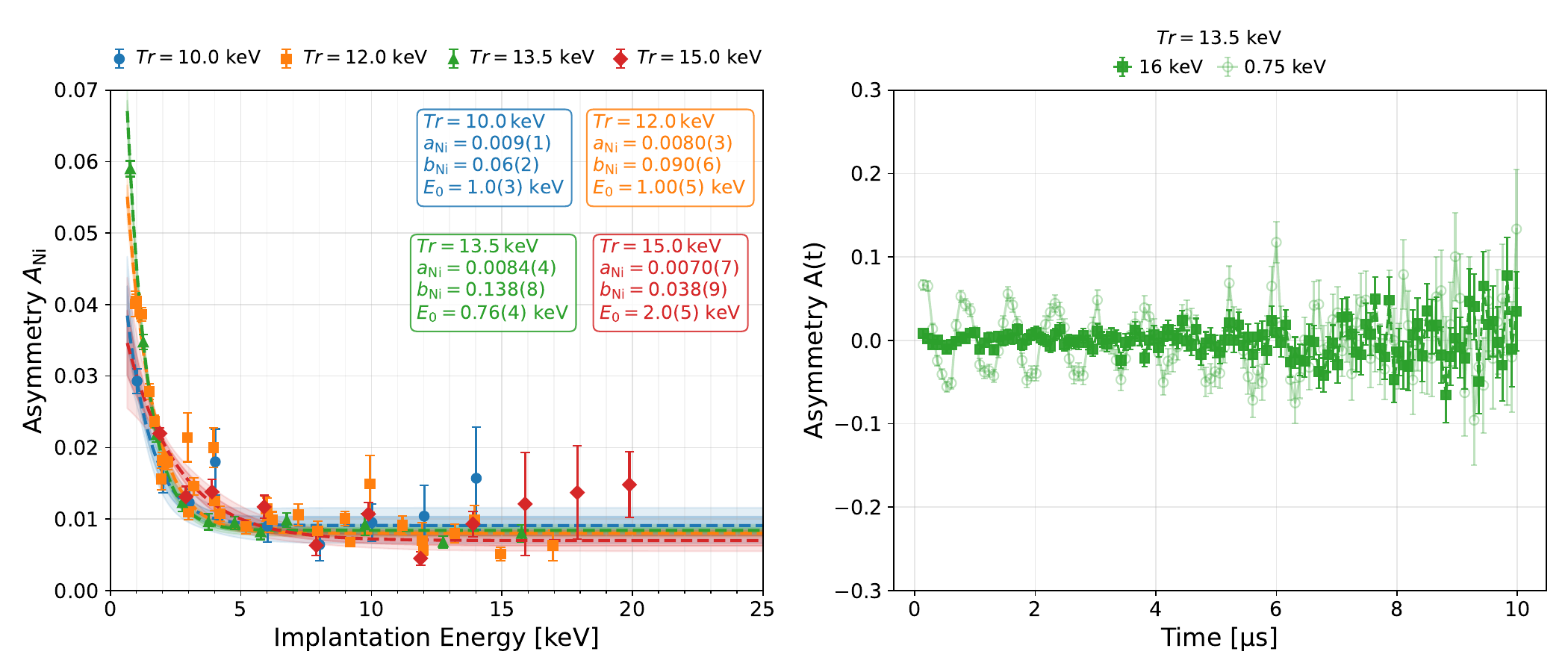}
\caption{\label{fig:reflection}
Left: Energy-dependent asymmetry $A_{\rm Ni}$ measured on a nickel-coated sample plate for transport settings of 10 (blue circles), 12 (orange squares), 13.5 (green triangles), and 15~keV (red diamonds). A pronounced increase in asymmetry is observed only at low implantation energies (typically below $\sim$ 2~keV), while at higher energies the asymmetry remains constant at a background level below 0.01. Dashed lines represent fits using Eq.~(\ref{eq:asyni}), the fit parameters are summarized in the top right corner. Right: Comparison of weak transverse-field $\mu$SR time spectra acquired at implantation energies of 0.75~keV (hollow circles) and 16~keV (full squares) for a transport setting of 13.5~keV, illustrating the appearance of a spurious oscillatory signal at low energies due to muons reflected at the sample plate and stopping in the surrounding radiation shield.
}
\end{figure*}
To obtain meaningful and reproducible calibration data, all measurements presented in the following were performed using a carbon foil that was cleaned by laser irradiation immediately before the measurements, as described in Ref.~\cite{2024_Janka, 2025_Janka}. Contamination of the carbon foil increases its effective thickness, leading to enhanced energy loss and energy straggling of the muon beam. This, in turn, modifies the neutralization probability at the foil and at the sample surface for a given transport setting. It also broadens the implantation-energy distribution and degrades the achievable beam spot at the sample position. Under such conditions, the depth- and transport-dependent asymmetry no longer reflects the nominal beam settings and cannot be reliably parameterized. Consequently, a clean carbon foil is an essential prerequisite for both accurate calibration measurements and for experiments that rely on quantitative analysis of depth-dependent asymmetry data.

The transport settings of 10, 12, 13.5, and 15~keV in the following measurements correspond to standard operating conditions of the LEM beamline. They were chosen to cover the range of implantation energies most commonly used in LE-$\mu$SR experiments. In particular, very low implantation energies (e.g.\ around 1~keV) are most stably achieved with a transport setting of 10~keV, whereas higher transport voltages (e.g., 15~keV) enable access to the largest implantation depths under stable high-voltage conditions on the sample.

\subsubsection{Silver Measurement}

To determine the maximally achievable transverse-field $A_0$, a silver-coated sample plate is mounted on the cryostat and the depth-dependent $A_0$ (here labeled as $A_{\rm Ag}$) is measured for different transport settings. Silver is non-magnetic and exhibits only negligible muon-spin depolarization; therefore, deviations of $A_{\rm Ag}$ from a constant value directly reflect instrumental effects.
In particular, these measurements capture $A_0$ losses due to backscattering from the sample surface and transport-dependent losses caused by muonium formation in the carbon foil. As such, $A_{\rm Ag}$ provides the most suitable normalization for depth-resolved measurements on samples that fully cover the muon beam spot, enabling, for example, the determination of magnetic volume fractions.

Fig.~\ref{fig:backscattering} shows the energy-dependent $A_{\rm Ag}$ for transport settings of 10~keV, 12~keV, 13.5~keV, and 15~keV. To parameterize the energy dependence in a form convenient for subsequent corrections, the data are fitted using a phenomenological model,
\begin{equation} \label{eq:asyag}
A_{\rm Ag}(E) = a_{\rm Ag} + b_{\rm Ag}\,\ln\!\left(\frac{E}{1\,\mathrm{keV}}\right),
\end{equation}
where $E$ denotes the implantation energy, while $a_{\rm Ag}$ and $b_{\rm Ag}$ are fit parameters representing the value of $A_{\rm Ag}$ at $E=1$~keV and the rate of increase in $A_{\rm Ag}$, respectively. The fitted parameters are included in the figure. The $\ln(E)$ dependence provides a good description of the data within the energy interval considered; however, it is known that $A_\mathrm{Ag}(E)$ would saturate at higher energies.

\begin{figure*}[t!]
\centering
\includegraphics[width=2\columnwidth, trim={0 0 0 0},clip]{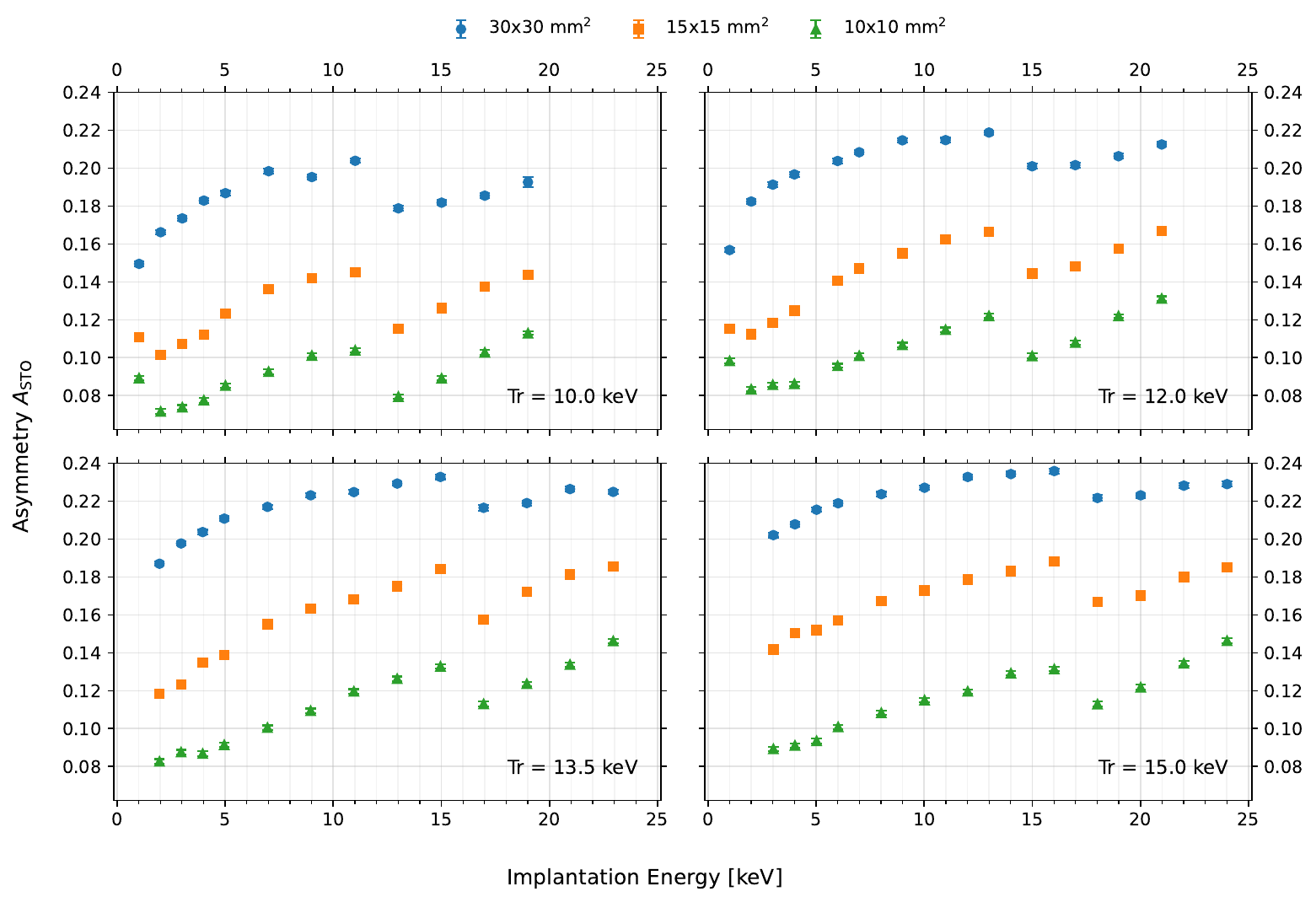}
\caption{\label{fig:STO_uncorr}
Energy-dependent wTF (100~G) asymmetry measured at 200~K on SrTiO$_3$ (STO) samples of different lateral dimensions (30~$\times$~30~mm$^2$ blue circles, 15~$\times$~15~mm$^2$ orange squares, and 10~$\times$~10~mm$^2$ green triangles) at various transport settings (10~keV top left, 12~keV top right, 13.5~keV bottom left, and 15~keV bottom right). As $f_\mathrm{dia}=100\%$ for STO measured at 200 K, deviations from a constant $A_\mathrm{STO}$ directly reflect instrumental effects discussed in the text.}
\end{figure*}

\subsubsection{Nickel Measurement}

The contribution from reflected muons is characterized independently using a nickel-coated sample plate. Nickel is ferromagnetic even at room temperature, and muons stopping in the Ni layer depolarize on timescales too fast to be resolved by the LEM spectrometer (within $\sim 0.07$~$\mu$s) \cite{Saadaoui2012PP}. Consequently, muons implanted into the Ni plate do not contribute a measurable TF asymmetry. Any observed asymmetry therefore originates from muons that are reflected at the sample plate and subsequently stop in the surrounding radiation shield.

Fig.~\ref{fig:reflection} (left) shows energy-dependent extracted values of $A_0$ (here labeled as $A_{\rm Ni}$) acquired for transport settings of 10~keV, 12~keV, 13.5~keV, and 15~keV. A significant increase in $A_{\rm Ni}$ is observed only at low implantation energies, typically below approximately 2~keV, while at higher energies $A_{\rm Ni}$ remains constant at a background level below 0.01. The data are well described using a phenomenological function
\begin{equation} \label{eq:asyni}
A_{\rm Ni}(E) = a_{\rm Ni} + b_{\rm Ni} \exp\!\left(-\frac{E}{E_0}\right),
\end{equation}
where $a_{\rm Ni}$ represents the flat background contribution, $b_{\rm Ni}$ the amplitude of the reflection-induced signal, and $E_0$ the characteristic energy scale over which reflection becomes relevant. The fitted parameters are included in the figure. Fig.~\ref{fig:reflection} (right) shows a direct comparison of the $\mu$SR time spectra at implantation energies of 0.75~keV and 16~keV for a transport setting of 13.5~keV, illustrating the appearance of a spurious oscillatory signal at low energies due to reflected muons.

\subsubsection{STO Measurement}

Fig.~\ref{fig:STO_uncorr} summarizes the combined impact of the instrumental effects discussed in Sec.~\ref{sec:effects}, using depth-resolved measurements on SrTiO$_3$ (STO), measured at $200$ K. Since $f_\mathrm{dia}=100\%$ in STO at temperatures above 70~K \cite{2014_ZaherSTO}, any deviation from a constant asymmetry directly reflects non-sample-related contributions. 

A systematic reduction of $A_0$ (here labeled as $A_\mathrm{STO}$) toward lower implantation energies is observed for all transport settings, consistent with increased muon backscattering at shallow depths. In addition, increases of $A_\mathrm{STO}$ at the lowest implantation energies are visible in some datasets, most notably for the 10~$\times$~10~mm$^2$ sample at Tr~=~10~keV, indicating contributions from muons reflected by the sample plate and stopping in the surrounding radiation shield.

The influence of the beam-sample overlap is evident from the comparison of different sample sizes. A progressive reduction of $A_\mathrm{STO}$ is observed when decreasing the sample dimensions from 30~$\times$~30~mm$^2$ to 15~$\times$~15~mm$^2$ and further to 10~$\times$~10~mm$^2$, demonstrating that an increasing fraction of muons misses the sample and lands on the Ni coated sample plate. 

At higher implantation energies, a sharp drop of $A_\mathrm{STO}$ is observed once the sample bias exceeds approximately $-3.6$~kV (e.g., at around 13 keV implantation energy for Tr~=~10~keV), consistent with the onset of secondary electron emission from the sample surface.

Finally, the maximum observable $A_\mathrm{STO}$ exhibits a pronounced dependence on the transport setting. For the largest STO sample, $A_\mathrm{STO}$ saturates slightly above 0.20 for Tr~=~10~keV, while it reaches approximately 0.24 for Tr~=~15~keV. This transport-dependent normalization directly reflects muon neutralization losses in the carbon foil and constitutes the dominant contribution to the transport dependence of the total asymmetry.

\subsection{\label{sec:sim}Simulation Input}

Within the \texttt{musrSim} framework \cite{musrSIM1, musrSIM2, Salman2012PP, 2015_Khaw, sedlak_musrsim_2012} based on \texttt{Geant4} \cite{Geant4}, the entire LEM beamline can be simulated. A new electrostatic field map of the sample environment was generated using \texttt{SIMION} \cite{Simion} and subsequently implemented in \texttt{musrSim}. In particular, the evolution of the muon beam spot as a function of the transport settings and the high voltage applied to the sample plate can be modeled and used to determine the implantation-energy-dependent overlap of the muon beam with a given sample area. 

The simulation was first tuned to reproduce the experimentally measured beam spots reported in Ref.~\cite{2024_Janka}. After this validation step, high-voltage scans of the sample plate were simulated, typically covering a range from $-14$~kV to $+14$~kV in steps of $1$~kV, and the corresponding beam spots at the sample position were recorded.

\begin{figure}[t!]
\centering
\includegraphics[width=\columnwidth, trim={0 0 0 0},clip]{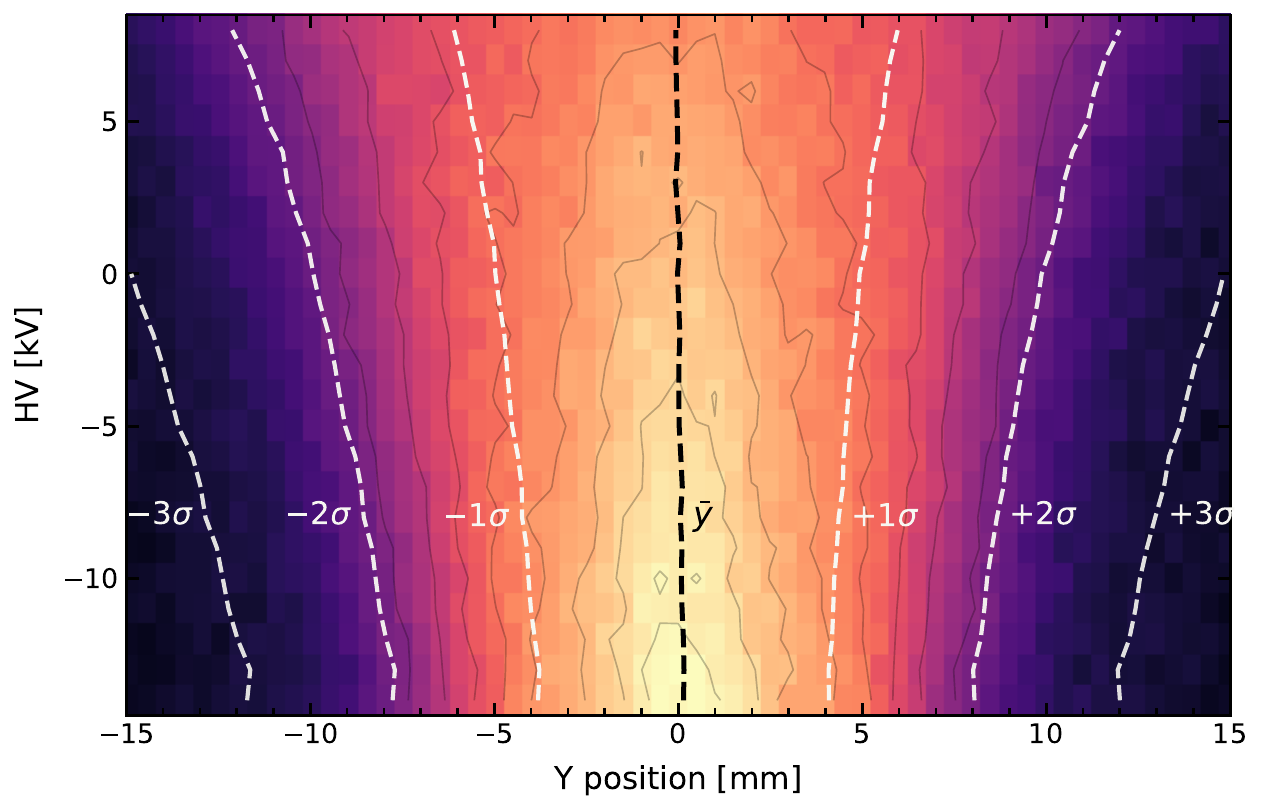}
\caption{\label{fig:beamspot_evolution}
Simulated evolution of the muon beam spot at the sample position for a transport setting of 13.5~keV as a function of the sample plate high voltage. For each HV value, the underlying two-dimensional beam distribution is projected onto the vertical ($y$) direction by integrating over $x$. The color scale represents the relative muon intensity (light colors indicate higher intensity). The dashed black line marks the mean vertical beam position $\bar{y}$, while the dashed white lines indicate the corresponding intensity-weighted RMS widths $\bar{y}\pm1\sigma$, $\bar{y}\pm2\sigma$, and $\bar{y}\pm3\sigma$. The voltage-dependent beam displacement and broadening determine the implantation-energy-dependent beam–sample overlap and form the basis of the sample-size correction factor.}
\end{figure}

From these simulations, the overlap between the muon beam and the sample area is calculated as a function of implantation energy. The samples have been assumed to be perfect squares (centered with the beam), since this is the most commonly used sample shape at LEM. The simulated overlap $O(E)$ is the fraction of muons stopping within the sample area; the remainder $1-O(E)$ misses the sample and is assumed to stop in the Ni-coated sample plate, contributing no transverse-field asymmetry, so that $O(E)$ serves as the corresponding geometrical correction factor. 
Fig.~\ref{fig:beamspot_evolution} shows the simulated evolution of the vertical beam profile at the sample position as a function of sample high voltage for $Tr=13.5$~keV; each high voltage corresponds to the $y$-projection of the two-dimensional beam spot (integrated over $x$). The mean vertical beam position $\bar{y}$ is indicated by a dashed black line, while the corresponding $\sigma$ bands (intensity-weighted RMS) are shown as dashed white lines. The color scale represents the relative intensity of the muon distribution, ranging from high intensity (white) to low intensity (black).
\begin{figure}[t!]
\centering
\includegraphics[width=1\columnwidth, trim={0 0 0 0},clip]{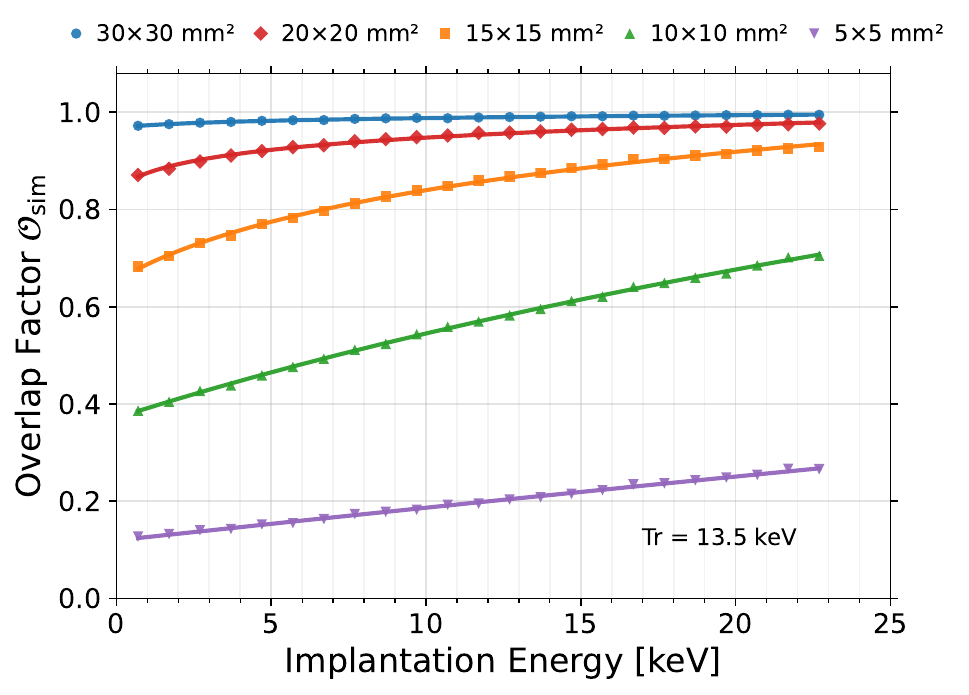}
\caption{\label{fig:simulation}
Simulated beam--sample overlap factor $\mathcal{O}_{\mathrm{sim}}$ as a function of implantation energy for different square sample sizes for 13.5~keV transport settings. Colored symbols denote different sample dimensions as indicated in the legend. Solid lines represent empirical fits used solely to parameterize the implantation-energy dependence of the overlap factor for interpolation and correction purposes. The resulting parameterization quantifies the fraction of muons reaching the sample and forms the basis for the sample-size-dependent asymmetry correction. The simulations assume perfectly centered samples.
}

\end{figure}

\begin{table}[t!]
\centering
\resizebox{\columnwidth}{!}{%
\begin{tabular}{c c c c c}
\toprule
Size [mm$\times$mm] & Tr [keV] & $\mathcal{O}_{\mathrm{sim}}(0)$ & $m [\mathrm{keV}^{-1}]$ & $E_0$ [keV] \\
\midrule
5$\times$5  & 10   & 0.093(2)  & 0.0073(5) & 200(286) \\
5$\times$5  & 12   & 0.112(2)  & 0.0073(4) & 181(181) \\
5$\times$5  & 13.5 & 0.119(2)  & 0.0069(3) & 200(181) \\
5$\times$5  & 15   & 0.133(2)  & 0.0066(3) & 200(161) \\
\midrule
10$\times$10 & 10   & 0.299(3)  & 0.0200(9) & 58(18) \\
10$\times$10 & 12   & 0.355(3)  & 0.0203(8) & 34(6) \\
10$\times$10 & 13.5 & 0.371(2)  & 0.0205(6) & 27(3) \\
10$\times$10 & 15   & 0.403(2)  & 0.0195(6) & 26(3) \\
\midrule
15$\times$15 & 10   & 0.536(7)  & 0.049(4)  & 4.6(7) \\
15$\times$15 & 12   & 0.626(5)  & 0.040(3)  & 4.1(5) \\
15$\times$15 & 13.5 & 0.654(4)  & 0.038(3)  & 3.8(4) \\
15$\times$15 & 15   & 0.684(6)  & 0.041(4)  & 2.8(4) \\
\midrule
20$\times$20 & 10   & 0.70(6)   & 0.3(3)    & 0.2(3) \\
20$\times$20 & 12   & 0.80(2)   & 0.13(7)   & 0.3(2) \\
20$\times$20 & 13.5 & 0.841(7)  & 0.05(1)   & 0.8(2) \\
20$\times$20 & 15   & 0.86(1)   & 0.06(3)   & 0.5(3) \\
\midrule
30$\times$30 & 10   & 0.949(8)  & 0.04(4)   & 0.3(3) \\
30$\times$30 & 12   & 0.967(1)  & 0.005(1)  & 2.5(8) \\
30$\times$30 & 13.5 & 0.9684(8) & 0.0067(10)& 1.4(3) \\
30$\times$30 & 15   & 0.9743(8) & 0.0033(6) & 3.1(8) \\
\bottomrule
\end{tabular}
}
\caption{\label{tab:overlap}Fit parameters $\mathcal{O}_{\mathrm{sim}}(0)$, $m$, and $E_0$ obtained from the beam--sample overlap simulations for different sample sizes and transport settings. The values are applicable to perfectly centered samples.}
\end{table}

Typically, $5\times10^{4}$ muons were simulated for each combination of transport setting and sample bias to determine the beam spot and beam--sample overlap. This number was sufficient to ensure that the statistical uncertainty of the overlap correction is negligible compared to the experimental uncertainties.

Fig.~\ref{fig:simulation} shows an example of the simulated beam--sample overlap factor $\mathcal{O}_{\mathrm{sim}}$ for a transport setting of 13.5~keV as a function of implantation energy for perfectly centered samples. 
To facilitate interpolation and a compact representation of the simulation results, the data were fitted with the empirical function
\begin{equation}
\mathcal{O}_{\mathrm{sim}}(E)
= \mathcal{O}_{\mathrm{sim}}(E=0) \;+\; m\,E_0\,\ln\!\left(1+\frac{E}{E_0}\right),
\end{equation}
which serves solely as a convenient parameterization of the energy dependence. The fit parameters are used to enable a consistent and reproducible overlap correction of experimental data across different implantation energies, sample sizes, and transport settings. The parameters are summarized in Table~\ref{tab:overlap}.

\section{\label{sec:results}Benchmark}

Taking into account the experimental as well as simulation inputs, the determination of the diamagnetic fraction described by Eq.~\ref{eq:diaE_full} becomes:
\begin{equation}
\label{eq:dia}
f_{\mathrm{dia}}(E)
=
\frac{A_{\mathrm{STO}}(E)-A_{\mathrm{Ni}}(E)}
{\mathcal{O}(E)\,\bigl[A_{\mathrm{Ag}}(E)-A_{\mathrm{Ni}}(E)\bigr]}.
\end{equation}
For the reference measurement at 200~K, where $f_{\mathrm{dia}}(E)=1$, Eq.~\ref{eq:dia} can be rewritten to extract the overlap factor $\mathcal{O}(E)$ directly from the measured asymmetries:
\begin{equation}
\mathcal{O}(E)
=
\frac{A_{\mathrm{STO}}(E)-A_{\mathrm{Ni}}(E)}
     {A_{\mathrm{Ag}}(E)-A_{\mathrm{Ni}}(E)}.
\end{equation}
This experimentally determined overlap factor can be directly compared to the simulation results described in Sec.~\ref{sec:sim}.

\begin{figure*}[t!]
\centering
\includegraphics[width=2\columnwidth, trim={0 0 0 0},clip]{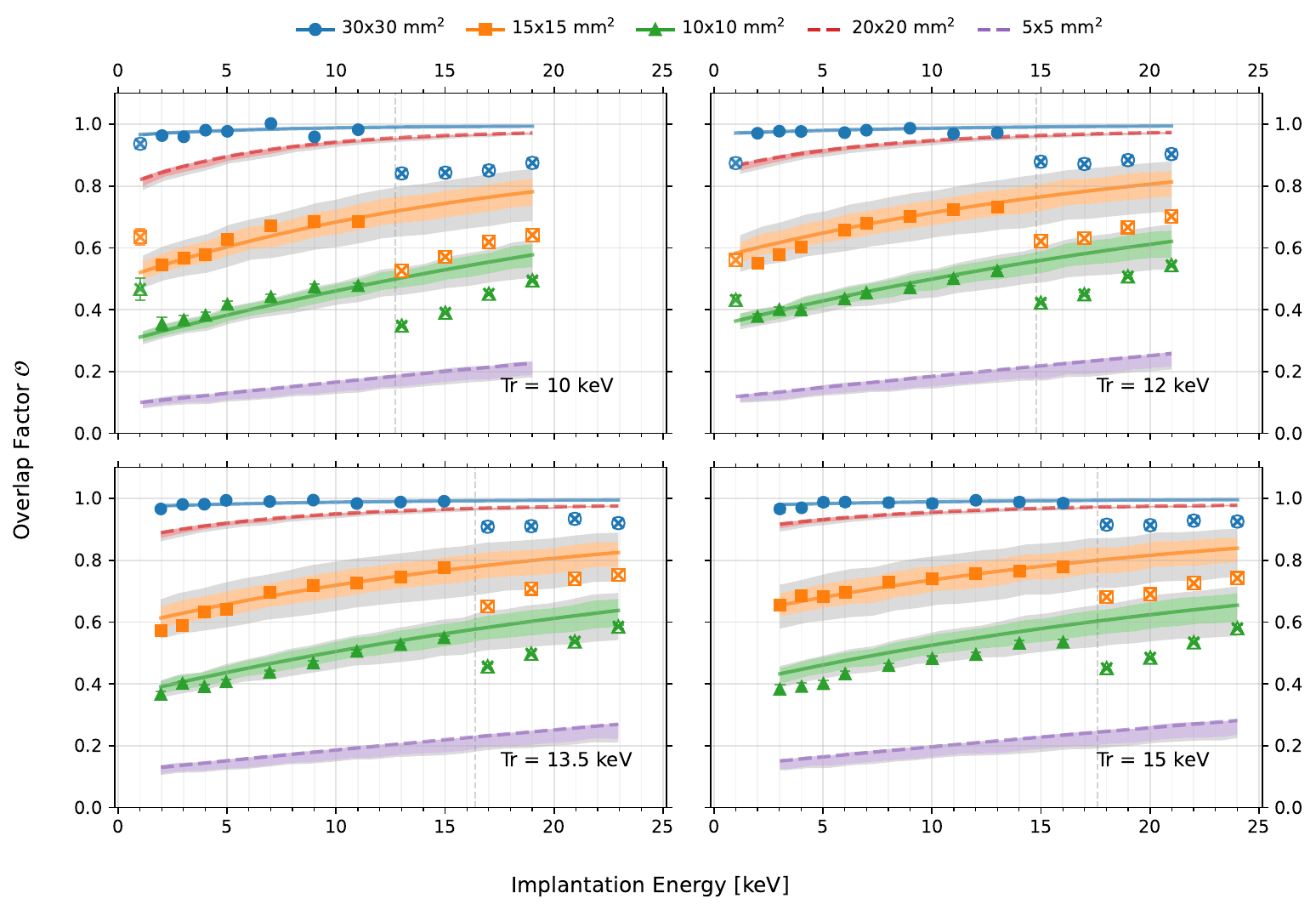}
\caption{\label{fig:STO_corr}
Benchmark comparison between experimentally determined overlap factors from SrTiO$_3$ (STO) measurements and simulated beam--sample overlap factors. Experimental data points are shown for sample sizes of 10$\times$10 (green triangles), 15$\times$15 (orange squares), and 30$\times$30~mm$^2$ (blue circles), where overlap factors were extracted from depth-resolved weak transverse-field STO data after normalization with Ag and correction for reflection using Ni reference measurements. Solid lines show the corresponding simulated overlap factors $\mathcal{O}_{\mathrm{sim}}(E)$ for these sample sizes. For sample sizes of 5$\times$5 and 20$\times$20~mm$^2$, no experimental STO data are available; the corresponding simulated overlap factors are therefore shown as dashed lines.
Shaded bands illustrate the sensitivity of the simulated overlap to sample misalignment: the colored bands correspond to the maximum expected uncertainty for beam--sample center offset of 0.5~mm, while the gray bands indicate a conservative estimate assuming a 1~mm offset. Data points marked with a cross ($\times$) indicate measurements affected by known systematic effects that are not included in the simulation (e.g. secondary electron emission) and are therefore not expected to be described by the overlap model. The four panels correspond to transport energies of 10~keV (top left), 12~keV (top right), 13.5~keV (bottom left), and 15~keV (bottom right).
}
\end{figure*}

For each measured sample, a photograph was taken to determine the lateral position of the STO sample relative to the beam axis. This information is incorporated into the simulations to account for possible sample misalignment. For the 10$\times$10~mm$^2$ sample, the required correction was $-1.5$~mm in the horizontal ($x$) direction and $-1.9$~mm in the vertical ($y$) direction. For the 15$\times$15~mm$^2$ sample, the corresponding offsets were $+1.2$~mm and $-2.6$~mm. In contrast, the 30$\times$30~mm$^2$ sample could be mounted more reliably using four 15$\times$15~mm$^2$ sample pieces, resulting in much smaller offsets of $+0.2$~mm and $+0.5$~mm, respectively. The resulting benchmark comparison between experiment and simulation is shown in Fig.~\ref{fig:STO_corr}. The simulated curves shown in Fig.~\ref{fig:STO_corr} are based on the same model as in Fig.~\ref{fig:simulation}, but include corrections for the experimentally determined sample misalignment.
The shaded bands in Fig.~\ref{fig:STO_corr} illustrate the sensitivity of the overlap correction to uncertainties in the relative positioning of the beam and the sample. Prior to each measurement campaign, the beam spot parameters are determined using a position-sensitive MCP installed at the sample position \cite{2024_Janka}. The beamline is tuned such that the beam spot deviates by no more than 0.5~mm from the nominal center. The colored bands therefore correspond to a relative beam--sample offset of 0.5~mm, while the gray bands represent a conservative estimate assuming a 1~mm misalignment. For sample sizes of 5$\times$5 and 20$\times$20~mm$^2$, no experimental STO data are available; only the simulated overlap factors with the corresponding misalignment bands are shown. For perfectly centered samples, the uncertainty bands extend only toward lower overlap factors, as illustrated for the 5$\times$5 and 20$\times$20~mm$^2$ cases. When correcting for sample misalignment, the uncertainty can also extend toward larger overlap factors, as observed for the experimentally benchmarked 10$\times$10 and 15$\times$15~mm$^2$ samples.

Overall, the comparison between experimental data and simulation demonstrates that, over a wide range of implantation energies and for all investigated transport settings, the simulation reproduces the experimentally extracted overlap factors with good accuracy. Deviations from the simulation are observed at implantation energies corresponding to sample bias voltages below $-3.6$~kV. In this bias regime, the dominant contribution arises from secondary electron emission at the sample surface, as discussed in Sec.~\ref{sec:electron_emission}. Because this effect depends sensitively on sample-specific electronic and surface properties (e.g., the secondary electron yield per implanted muon), it cannot be accounted for by a universal correction based on reference measurements alone. Consequently, it is not included in the present correction framework, and the affected data points are not described by the overlap model.

Additional deviations are observed at the lowest implantation energy (1~keV). Whether these originate from instrumental effects, from a thin contaminated surface layer, or from variations in backscattering between different materials cannot be conclusively determined within the present study.

From these results, it is evident that small sample sizes, such as 5$\times$5 and 10$\times$10~mm$^2$, exhibit a strongly reduced beam--sample overlap under the present beam conditions, leading to a substantial reduction of the measurable asymmetry. Consequently, higher counting statistics are required to achieve comparable precision, resulting in longer measurement times. Although the overlap correction developed here allows this effect to be quantified, the associated uncertainty increases markedly as the overlap decreases. Quantitative depth-resolved analysis therefore becomes progressively less robust for smaller samples, and larger sample dimensions remain preferable for LE-$\mu$SR studies under standard beam conditions.

The recommended sample size for LE-$\mu$SR is 20$\times$20~mm$^2$ or larger; however, the present analysis shows that even for 20$\times$20~mm$^2$ samples, up to 20\% of the asymmetry can be lost. Importantly, this loss can be reliably corrected using the simulation-based overlap correction, introducing only a modest additional uncertainty.

If the depth-dependent asymmetry is a key quantity to be extracted from LE-$\mu$SR measurements, careful mounting of the sample on the sample plate is essential. The lateral positioning accuracy should be better than 1~mm. Increasing the sample size further reduces the sensitivity of the extracted asymmetry to residual beam--sample misalignment and improves the robustness of the correction procedure.

When small samples (5$\times$5 to 10$\times$10~mm$^2$) cannot be avoided, the installation of an upstream collimator can reduce the beam spot and improve the geometric overlap \cite{2023_Ni}. This mitigation strategy, however, reduces the available muon flux significantly and therefore requires longer acquisition times to reach comparable statistical precision.

\section{Summary and Outlook}

We have established an updated calibration procedure for the energy-dependent total asymmetry in low-energy $\mu$SR at the LEM beamline under current operating conditions. Updated silver and nickel reference measurements account for transport- and energy-dependent asymmetry losses and spurious contributions following the installation of a new carbon foil in the single-muon tagging system.

In addition, we introduced a simulation-based correction for sample-size-dependent beam--sample overlap using detailed \texttt{Geant4}-based LEM beamline simulations with an updated electrostatic field map. The resulting overlap factors were parameterized for practical use and benchmarked experimentally using depth-resolved weak transverse-field measurements on SrTiO$_3$. The good agreement between experiment and simulation validates this approach over a wide range of implantation energies and transport settings.

Together, these corrections enable a reliable extraction of depth-dependent magnetic volume and muonium formation fractions in LE-$\mu$SR experiments. The analysis also highlights that the uncertainty associated with the asymmetry correction increases rapidly as the beam--sample overlap decreases. While the correction framework allows measurements on smaller samples to be interpreted, such measurements are intrinsically less robust and require particularly careful sample mounting and accurate centering. Larger sample sizes therefore remain strongly preferable for quantitative LE-$\mu$SR studies. The presented framework provides a robust and flexible basis for quantitative LE-$\mu$SR analysis and can be readily adapted to future beamline upgrades, modified sample environments, or the introduction of new corrections, such as sample-dependent secondary electron emission.

\begin{acknowledgments}
All the measurements have been performed at the Swiss Muon Source S$\mu$S, Paul Scherrer Institute, Villigen, Switzerland.
\end{acknowledgments}

\section*{Data Availability Statement}

The $\mu$SR data that support the findings of this study are openly available in the SciCat PSI repository at http://doi.psi.ch/detail/10.16907/901d6fda-db9a-4925-992a-4a9a9410c72d.

\nocite{*}
\bibliography{aipsamp}
\bibliographystyle{apsrev4-1}

\end{document}